\documentclass[a4paper]{article}
\begin{document}

\title{Regularization of a three-body problem with zero-range potentials}

\author{D.V.Fedorov and A.S.Jensen \\
IFA, {\AA}rhus University, 8000 {\AA}rhus C, Denmark}

\date{}

\maketitle

\begin{abstract}
We propose a coordinate-space regularization of the three-body problem
with zero-range potentials.  We include the effective range and the shape
parameter in the boundary condition of the zero-range potential. The
proposed extended zero-range model is tested against atomic helium trimers
and is shown to provide an adequate quantitative description of these systems.
\end{abstract}

\section{Introduction}

The zero-range potential \cite{demkov} has been extensively used
over many years as a practically convenient form of the effective
interaction. The concept employs separation of scales in a physical
problem and allows qualitative and often quantitative description of
low-energy properties of a physical system in a simple and transparent
way (see, e.g., \cite{jackson}).

However, application of the zero-range potential to a three-body system
presents a problem -- a {\em collapse} of the system known as the Thomas
effect \cite{thomas}. The three-body system with zero-range potentials
has no ground state but infinitely many bound states with vanishing
spatial extension and exceedingly large binding energy.

Several attempts have been made to alleviate this problem by
adding some sort of cutoff to the potential in momentum space
\cite{esbensen,amorim,bedaque} or by switching to a finite-range
potential in certain areas of configuration space \cite{esben-macek}.

We introduce an alternative coordinate space approach where the collapse
is removed by a suitable modification of the boundary condition of the
zero-range potential.  The boundary condition is extended to include the
higher order parameters of the effective range expansion.  The three-body
system acquires then a well defined ground state while all the simplicity
and transparency of the zero-range model is retained.

We apply the developed formalism to rather involved three-body systems,
helium trimers, and show that the extended zero-range model provides
an accurate description of these systems.

\section{The zero-range model and regularization}

\paragraph{Zero-range potentials.}
The quantum mechanical two-body problem with a zero-range potential can
be formulated \cite{demkov} as the free Schr\"{o}dinger equation for the
s-state wave-function $\psi$ with the relative coordinate $r$ and wave
number $k$,
\begin{equation}
\left( -\frac{d^{2}}{dr^{2}}-k^{2}\right) r\psi=0 \;,
\end{equation}
with the solution 
\begin{equation}
r\psi=\sin \left( kr+\delta (k)\right) \;,
\end{equation}
and a boundary condition at $r=0$ expressed in terms of the scattering
length $a$ as
\begin{equation}\label{2bbc}
\left. \frac{1}{r\psi}\frac{d(r\psi)}{d r} \right|
_{r=0}=k\cot \delta (k)=\frac{1}{a}\;.
\end{equation}

For negative scattering length a bound state solution exists
\begin{equation}
r\psi\propto \exp (-\kappa r)\;,
\end{equation}
where $\kappa >0$ can be found from the boundary condition (\ref{2bbc}),
$\kappa=1/|a|$.

The zero-range model for a three-body system can be formulated as a
{\em free} three-body wave-function $\Psi $ which satisfies the three
boundary conditions
\begin{equation}
\left. \frac{1}{\left| {\bf r}_{j}-{\bf r}_{k}\right| \Psi 
}\frac{\partial
\left| {\bf r}_{j}-{\bf r}_{k}\right| \Psi }{\partial \left| {\bf r}_{j}-
{\bf r}_{k}\right| }\right| _{\left| {\bf r}_{j}-{\bf r}_{k}\right| =0}=
\frac{1}{a_{i}}\;,\;i=1,2,3\;,  \label{3bc}
\end{equation}
where ${\bf r}_{i}$ is the coordinate of $i$-th particle, $a_{i}$ is
the scattering length in the two-body system of particles $j$ and $k$
with $\{i,j,k\}$ being a positive permutation of $\{1,2,3\}$.

The derivatives in the boundary condition (\ref{3bc}) are most suitably
formulated in terms of the hyper-spheric coordinates $\{\rho ,\alpha_i\}$
(defined in the appendix):
\begin{equation}
\left. \frac{\partial }{\partial \left| {\bf r}_{j}-{\bf r}_{k}\right| }
\right| _{\left| {\bf r}_{j}-{\bf r}_{k}\right| =0}=\frac{\sqrt{\mu _{i}}}{
\rho }\left. \frac{\partial }{\partial \alpha _{i}}\right| _{\alpha
_{i}=0}\;.
\end{equation}
The boundary condition (\ref{3bc}) can then be rewritten as 
\begin{equation}\label{3bc2}
\left.\frac{1}{\alpha _{i}\Psi }\frac{\partial(\alpha _{i}\Psi)}{\partial
\alpha _{i}}\right|_{\alpha _{i}=0}=\frac{\rho }{\sqrt{\mu _{i}}}\frac{1}{
a_{i}}\;.
\end{equation}

\paragraph{Hyper-spheric expansion.}
We shall employ the hyper-spheric adiabatic expansion \cite{review}
of the three-body wave-function
\begin{equation}\label{expan}
\Psi (\rho ,\Omega )=\frac{1}{\rho ^{5/2}}\sum_{n}f_{n}(\rho )\Phi _{n}(\rho
,\Omega )\;,
\end{equation}
in terms of the complete basis $\Phi_n(\rho,\Omega)$ of the solutions of
the hyper-angular eigenvalue equation
\begin{equation}\label{eigen}
\left( \Lambda+\frac{2m\rho ^{2}}{\hbar ^{2}}
\sum_{i=1}^{3}V_{i}\right) \Phi_n(\rho ,\Omega )
=\lambda _{n}(\rho )\Phi_n(\rho,\Omega ) \;,
\end{equation}
where $V_{i}$ is the potential between particles $j$ and $k$, $m$ is
the mass scale used in the definition of the hyper-spheric coordinates
and $\Lambda$ is the angular part\footnote{Here $\Omega$ is any of the
tree possible sets of angles} of the kinetic energy operator (see
appendix).

For potentials without strong repulsive cores already the lowest term in the
expansion -- the so called hyper-spheric adiabatic approximation -- gives a
very good approximation to the precise solution \cite{esbenJPB}.
Again it is the lowest term that causes the Thomas collapse of a
three-body system with zero-range potentials. Therefore in the following
for the sake of simplicity we shall consider only this problematic
lowest term of the hyper-spheric expansion. Inclusion of higher terms
is straightforward.

The wave-function then simplifies to 
\begin{equation}\label{adia}
\Psi (\rho ,\Omega )=\frac{1}{\rho ^{5/2}}f(\rho )\Phi (\rho ,\Omega )\;,
\end{equation}
where the hyper-radial wave-function $f(\rho)$ satisfies the equation 
\begin{equation}\label{hyprad}
\left( -\frac{\partial^2}{\partial\rho^2}+\frac{\lambda(\rho)
+\frac{15}{4}}{
\rho ^{2}}-Q(\rho )-\frac{2mE}{\hbar ^{2}}\right) f(\rho ) =0\;,
\end{equation}
where $\lambda(\rho)$ is the lowest eigenvalue in eq.~(\ref{eigen}), $E$
is the total energy and
\begin{equation}
Q(\rho ) =\int d\Omega \Phi (\rho ,\Omega )\frac{\partial ^{2}}{\partial
\rho ^{2}}\Phi (\rho ,\Omega )\;.
\end{equation}

\paragraph{Faddeev equations.}

For short-range and zero-range potentials the Faddeev decomposition
of the angular wave-function $\Phi(\rho,\Omega)$ provides a convenient
framework for an analysis of the three-body system \cite{fed93},
\begin{equation}\label{dec}
\Phi(\rho,\Omega)=
\sum_{i=1}^{3}\frac{\varphi_i(\rho,\alpha_i)}{\sin(2\alpha_i)},
\end{equation}
where the three components $\varphi _{i}(\rho,\alpha _{i})$ satisfy a
system of Faddeev equations \cite{faddeev} 

\begin{equation}\label{fad}
\left( \Lambda-\lambda(\rho) \right)
\frac{\varphi_i(\rho,\alpha_i)}{\sin (2\alpha _{i})}
+ \frac{2m\rho ^{2}}{\hbar ^{2}}V_{i}\Phi(\rho,\Omega)=0\;.
\end{equation}
Since the zero-range potentials act only on the $s$-waves only the latter
are included in each of the three components $\varphi _{i}$.

All three components of the wave-function $\Phi$ in (\ref{fad}) must be
``rotated'' into the same Jacobi system. This is done by substituting
the variables and subsequently projecting onto the s-waves.  The
transformation of $\varphi_k$ into the $j$'th Jacobi system is given as
\begin{equation}
\varphi _{j\leftarrow k}(\alpha _{j})=\frac{1}{\sin (2\phi _{jk})}
\int_{\left| \phi _{jk}-\alpha _{j}\right| }^{\frac{\pi }{2}-\left|
\frac{\pi}{2}-\phi_{jk}-\alpha _{j}\right| }
\varphi _{k}(\alpha _{k})d\alpha_{k},
\end{equation}
where 
\begin{equation}
\phi _{jk}=\arctan \left( \sqrt{\frac{m_{i}(m_{1}+m_{2}+m_{3})}{m_{j}m_{k}}}
\right) .
\end{equation}
The expansion of $\varphi_{j\leftarrow k}(\alpha_{j})$ for small angles
$\alpha_j\ll~1$ reads
\begin{equation} \label{rot2} 
\varphi _{j\leftarrow k}(\alpha _{j})=\alpha _{j}\frac{2\varphi _{k}(\phi
_{jk})}{\sin (2\phi _{jk})}+O(\alpha _{j}^{2}) \;.
\end{equation}

The zero-range potentials vanish identically except at the origin and
we are therefore left with the free Faddeev equations
\begin{equation}
\left( -\frac{\partial ^{2}}{\partial \alpha _{i}^{2}}-\nu^{2}(\rho
)\right) \varphi _{i}(\rho ,\alpha _{i}) =0\;,
\end{equation}
which are obtained from (\ref{fad}) with $l_x=l_y=0$ and $V_i$=0, and
where $\nu ^{2}=\lambda +4$.
The solutions are
\begin{equation}\label{free}
\varphi _{i}(\rho ,\alpha _{i}) =A_{i}(\rho )\sin \left[ \nu (\rho )\left(
\alpha _{i}-\frac{\pi }{2}\right) \right]
\end{equation}
with the boundary condition $\phi_{i}(\rho,\frac{\pi}{2})=0$.

The factors $A_{i}$ are to be determined from the boundary condition
(\ref{3bc2}) which can now be reformulated in terms of the angular function
$\Phi$ as
\begin{equation}\label{bcnew}
\left.
\frac{\partial\left(\alpha_i\Phi(\rho,\Omega)\right)}{\partial\alpha_i}
\right|_{\alpha_i=0}=
\left.
\frac{\rho}{\sqrt{\mu_i}}\frac{1}{a_i}
\alpha_i\Phi(\rho,\Omega)
\right|_{\alpha_i=0} \;.
\end{equation}

The needed wave-function $\alpha_i\Phi$ and its partial derivative
$\partial(\alpha_i\Phi)/\partial\alpha_i$ are easily obtained
from (\ref{dec}) and (\ref{rot2})
\begin{eqnarray}
2\alpha_i\Phi =\varphi_i(\alpha_i)+
\alpha_i\sum_{j\ne i}\frac{2\varphi_j(\phi_{ij})}{\sin(2\phi_{ij})}
+O(\alpha_j^{2}) \;,\\
\left. 2\frac{\partial(\alpha_i\Phi)}{\partial\alpha_i}
\right|_{\alpha_i=0}=
\left. \frac{\partial\varphi_i(\alpha_i)}{\partial\alpha_i}
\right|_{\alpha_i=0}+
\sum_{j\ne i}\frac{2\varphi_{j}(\phi_{ij})}{\sin(2\phi_{ij})} \;.
\end{eqnarray}

Substituting the free solutions (\ref{free}) leads to
\begin{eqnarray}
\left. 2\alpha_i\Phi \right|_{\alpha_i=0}&=&
-A_{i}\sin \left( \nu \frac{ \pi }{2}\right) \;, \\
\left. 2\frac{\partial(\alpha_{i}\Phi)}{\partial \alpha _{i}}\right|
_{\alpha _{i}=0}&=&
A_{i}\nu \cos \left( \nu \frac{\pi }{2}\right)
+\sum_{j\ne i}A_{j}\frac{
2\sin \left[ \nu \left( \phi _{ij}-\frac{\pi }{2}\right) \right] }{\sin
(2\phi _{ij})}
\end{eqnarray}

The boundary condition (\ref{bcnew}) then becomes a
system of linear equations for the three factors $A_{i}$ 
\begin{equation}
A_{i}\nu \cos \left( \nu \frac{\pi }{2}\right) +\sum_{j\ne i}A_{j}\frac{
2\sin \left[ \nu \left( \phi _{ij}-\frac{\pi }{2}\right) \right] }{\sin
(2\phi _{ij})}= -
\frac{\rho }{\sqrt{\mu_i}}\frac{1}{a_{i}}
A_{i}\sin \left( \nu \frac{\pi }{2}\right)
\;. \end{equation}
A non-trivial solution exists only when the determinant of the
corresponding matrix $M(\nu,\rho)$ is zero
\begin{equation}\label{det}
\det M(\nu,\rho)=0\;,
\end{equation}
where the matrix elements are
\begin{eqnarray}
M_{ii} &=&
\nu \cos \left( \nu \frac{\pi }{2}\right) +\sin \left( \nu \frac{
\pi }{2}\right) \frac{\rho }{\sqrt{\mu _{i}}}\frac{1}{a_{i}}\;, \\
M_{i\neq j} &=&
\frac{2\sin \left[ \nu \left( \phi _{ij}-\frac{\pi }{2}
\right) \right] }{\sin (2\phi _{ij})}\;.  \nonumber
\end{eqnarray}
The solution $\nu (\rho )$ of the equation (\ref{det}) provides the
adiabatic potential $(\nu^2(\rho)-1/4)/\rho ^{2}$ for the hyper-radial
equation (\ref{hyprad}) from which one obtains the hyper-radial
wave-function of the three-body system.

\paragraph{Asymptotic behaviour of the eigenvalues.}
For a system of three identical bosons, where $\varphi_{ij}=\pi/3$,
equation (\ref{det}) simplifies to
\begin{equation}\label{efim}
\frac{-\nu \cos (\nu \frac{\pi }{2})+\frac{8}{\sqrt{3}}
\sin (\nu \frac{\pi }{ 6})}{\sin (\nu \frac{\pi }{2})}=
\frac{\rho }{\sqrt{\mu }}\frac{1}{a}\;.
\end{equation}
For large distances, $\rho\gg a$, there is a solution
that asymptotically approaches $\nu(\infty)=2$. Expanding (\ref{efim})
in terms of $1/\rho$ around $\nu$=2 gives the leading terms
\begin{equation}
\nu =2-\frac{12}{\pi }\frac{\sqrt{\mu }a}{\rho }\;,\;\frac{\lambda }{\rho
^{2}}=-\frac{16}{\pi }\frac{3\sqrt{\mu }a}{\rho ^{3}}\;,
\end{equation}
which is the lowest solution when no bound two-body subsystems are
present.  In this case the effective potential is of $1/\rho^3$ type.

However, when there is a two-body bound state another kind of solution
exists for large $\rho $ which asymptotically behaves as $\nu\sim i\rho
$. The leading terms then are
\begin{eqnarray}
\nu &=&i\frac{\rho }{\sqrt{\mu }}\frac{1}{\left| a\right| }
+i\frac{8}{\sqrt{3}}
\exp(-\frac{\rho}{\sqrt{\mu}}\frac{1}{\left|a\right|}\frac{\pi}{3})\;,
\\
\lambda &=&-\frac{\rho ^{2}}{\mu a^{2}}-\frac{\rho }{\sqrt{\mu }}\frac{1}{
\left| a\right| }\frac{16}{\sqrt{3}}
\exp(-\frac{\rho }{\sqrt{\mu }}\frac{1}{\left|a\right|}\frac{\pi}{3})-4\;.
\end{eqnarray}
The effective potential is then of Yukawa type 
\begin{equation}
\frac{\lambda +15/4}{\rho ^{2}} =-\frac{2mB}{\hbar ^{2}}-\frac{1}{4\rho
^{2}}-\frac{16\sqrt{3}}{\pi }\frac{b}{\rho }
\exp(-\frac{\rho }{b})\;,
\end{equation}
where $B=\hbar^2/(2\mu ma^2)$ is the two-body binding energy and
$b=3\sqrt{\mu}\left|a\right|/\pi\;$.  The corresponding angular wave
function (\ref{free}) asymptotically is
\begin{equation}
\sin \left[ \nu \left( \alpha -\frac{\pi }{2}\right) \right] =\sin \left[ 
\frac{i\rho }{\sqrt{\mu }\left| a\right| }\left( \alpha -\frac{\pi }{2}
\right) \right]
\propto \exp \left( -\frac{\rho \alpha }{\sqrt{\mu }\left| a\right| }
\right) \;.
\end{equation}
This wave-function is non-vanishing only when
$\alpha\sim\sqrt{\mu}\left|a\right|/\rho\ll~1$.  In this region the Jacobi
coordinates $x$ and $y$ (defined in the appendix) are approximately,
up to the linear terms in $\alpha$, given by $x\approx\rho\alpha$ and
$y\approx\rho$. This solution corresponds to a bound two-body state
with the momentum $k_0=i/\left|a\right|$ and the binding energy
$B$. The three-body wave-function then factorizes as
\begin{equation}
\Phi \propto \frac{1}{x}\exp(-\frac{x}{\sqrt{\mu }\left| a\right| })f(y) 
\end{equation}
and describes a dimer in the bound state
$\frac{1}{x}\exp(-\frac{x}{\sqrt{\mu}\left|a\right|})$ and a third particle
with a relative coordinate $y$ and the wave-function $f(y)$. The
corresponding radial equation asymptotically describes a two-body
system with a Yukawa potential
\begin{equation}
\left[ -\frac{\partial ^{2}}{\partial \rho
^{2}}-\frac{2m}{\hbar ^{2}}(E+B)- \frac{16\sqrt{3}}{\pi }\frac{b}{\rho
}\exp(-\frac{\rho }{b})\right] f(\rho )=0\;.
\end{equation}
The term $-1/(4\rho^2)$ in this equation cancelled the leading order
term of $Q(\rho )$. Indeed the normalized angular Faddeev component is
(asymptotically)
\begin{equation}
\varphi (\rho ,\alpha )=\sqrt{\frac{2\rho }{\sqrt{\mu
}\left| a\right| }}\exp(-\rho \frac{\alpha }{\sqrt{\mu }\left|a\right|})
\end{equation}
and therefore
\begin{equation} Q(\rho )\rightarrow
\int_{0}^{\infty }\varphi (\rho ,\alpha )\frac{\partial ^{2}}{\partial
\rho ^{2}}\varphi (\rho ,\alpha )d\alpha =\allowbreak -\frac{1 }{4\rho
^{2}} \;.
\end{equation}
We have thus a correct asymptotic wave-function corresponding to a dimer
and a third particle in a relative $s$-wave.

The term $Q$ is generally small and only is important to ensure the
correct asymptotic behaviour. In the following practical application we
shall always for simplicity use only the leading term $-1/(4\rho^2)$
instead of the full $Q$ similar to the Langer correction term in
\cite{esben-macek}.

\paragraph{The Thomas effect and regularization.}

For $\rho \ll a$ the equation (\ref{efim}) for $\nu$ reduces  to
\begin{equation}
-\nu \cos (\nu \frac{\pi }{2})+\frac{8}{\sqrt{3}}\sin (\nu \frac{\pi }{6}
)=0\;,
\end{equation}
which has well known imaginary roots $\nu_{0}=\pm ig$, where
$g\cong 1.006$, which cause the Thomas and also the Efimov
\cite{efimov} effects.

These imaginary roots lead to an effective potential in
the hyper-radial equation which in the small distance region, $\rho
\ll a$, is equal to $(\nu _{0}^{2}-1/4)/\rho ^{2}\cong -1.262/\rho ^{2}$
and the radial equation becomes
\begin{equation}
\left( -\frac{\partial ^{2}}{\partial \rho ^{2}}+\frac{\nu _{0}^{2}-1/4}{
\rho ^{2}}-\frac{2mE}{\hbar ^{2}}\right) f(\rho )=0\;.
\end{equation}
The (negative) energy $E=-\hbar ^{2}\kappa ^{2}/(2m)$ is negligible compared
to the effective potential when the distance is sufficiently small, $\rho
\ll \kappa ^{-1}$, and the corresponding radial equation, 
\begin{equation}
\left( -\frac{\partial ^{2}}{\partial \rho ^{2}}+\frac{\nu _{0}^{2}-1/4}{
\rho ^{2}}\right) f(\rho )=0\;,
\end{equation}
has in this region solutions of the form $f(\rho )\sim \rho ^{n}$, where $n=
\frac{1}{2}\pm \nu _{0}$. For imaginary $\nu _{0}=\pm ig$ the exponent $n$
also acquires an imaginary part $\pm ig$ leading to 
\begin{equation}
f(\rho )\propto \sqrt{\rho }\exp({\pm ig\ln \rho })\;.  \label{logwav}
\end{equation}
This wave-function has infinitely many nodes at small distances
or, correspondingly, infinitely many low lying states at smaller
distances. This is called the Thomas effect.

A suitable modification of the boundary condition (\ref{3bc2}) is
necessary in order to eliminate the problematic imaginary root $\nu_0$
at $\rho=0$ which causes the Thomas effect.  Intuitively one could
generalize the zero-range potential by introducing the higher order
terms of the effective range theory,
\begin{equation}
\left. \frac{1}{r\psi}\frac{d(r\psi)}{dr}\right| _{r=0}
=\frac{1}{a}+\frac{1}{2}Rk^{2}+PR^{3}k^{4}\;,
\end{equation}
where $R$ is the effective range and $P$ is the shape parameter of the
two-body system.  This would lead to the following  modification of
the matrix elements in the eigenvalue equation (\ref{det})
\begin{eqnarray}\label{modm}
M_{ii} &=&
\nu\cos\left( \nu \frac{\pi }{2}\right)
+\sin\left(\nu\frac{\pi}{2}\right)\frac{\rho}{\sqrt{\mu_i}}\nonumber\\
&\times&\left[
\frac{1}{a_i}+\frac{1}{2}R_i\left(\frac{\sqrt{\mu_i}\nu}{\rho}\right)^2
+P_iR_i^3\left(\frac{\sqrt{\mu_i}\nu}{\rho}\right)^4
\right] \;,
\end{eqnarray}
and equation (\ref{3bc2}) for three identical bosons
is then replaced by an extended equation
\begin{equation}\label{extbc}
\frac{-\nu \cos (\nu \frac{\pi }{2})
+\frac{8}{\sqrt{3}}\sin (\nu \frac{\pi }{6})}{\sin(\nu\frac{\pi}{2})}
=\frac{\rho}{\sqrt{\mu}}
\left[
\frac{1}{a}+\frac{1}{2}R\left(\frac{\sqrt{\mu}\nu}{\rho}\right)^2
+PR^3\left(\frac{\sqrt{\mu}\nu}{\rho}\right)^4
\right] \;.
\end{equation}
This extended equation at $\rho=0$ has a real root $\nu(0)=0$ and the
Thomas collapse is therefore removed.

Although the second order term with the effective range is, in principle,
enough for elimination of the imaginary root, the fourth order term
is necessary to ensure the correct analytic properties of the roots of
the equation.

Unlike the scattering length and effective range the parameter $P$ has
to be interpreted as a regularization parameter which somehow accounts
for all the higher order terms in the $k^2$ expansion, rather than a
true shape parameter of the two-body scattering.  The scattering length
and effective range are important for the correct asymptotic behaviour
of the eigenvalue $\lambda$ at large distances while the $P$ parameter
accounts for the pocket region and is in this model supposed to absorb
all the remaining short distance properties of the system.

\section{Application to helium trimers}

\begin{figure}[t]
\begin{center} \input{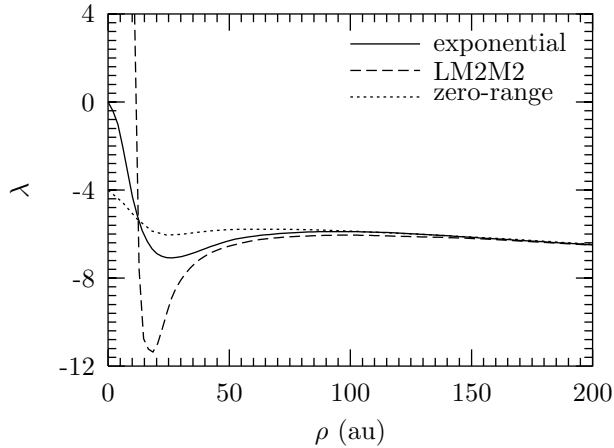} \end{center}
\caption{ The angular eigenvalue $\lambda$ as function of $\rho$
for the $^4$He-trimer for different potential models: exponential
\protect\cite{esbenJPB} , realistic LM2M2 \protect\cite{esbenJPB}, and
zero-range with $P$=0.13. All models have the same scattering length
$a$=-189.05~au and effective range $R$=13.843~au.}
\end{figure}

The helium trimer $^{4}$He$_{3}$ is a challenging three-body system
as there is a weakly bound dimer state, $^{4}$He$_{2}$, where the
scattering length $a$=-189.054~au is much larger than the effective
range $R$=13.843~au (the atomic unit of length is equal to the Bohr
radius $a_B$=0.529177~\AA).  Numeric computations with realistic
LM2M2 potential show that there is a ground state and an extremely
weakly bound excited state interpreted as an Efimov state (see, e.g.,
\cite{esbenJPB,sofianos} and references therein).

For our calculations we use as in \cite{esbenJPB} the mass scale
$m$=1822.887~au (the atomic unit of mass is equal to the electron mass
$m_e$=0.510999~MeV/$c^{2}$). The mass of the $^{4}$He atom is
$m$($^{4}$He)=4.002603$m$. The angular eigenvalue
$\lambda(\rho)=\nu^2(\rho)-4$ is obtained directly by numeric solution
of the transcendental equation (\ref{extbc}).

On Fig.~1 we compare the angular eigenvalues obtained from the zero-range
model with $P$=0.13 and from two finite-range models: the realistic
LM2M2 potential, and an exponential potential -- all models having the
same scattering length and effective range.

At large distances, $\rho\gg R$, the angular eigenvalues from all models
approach each other since it is only the scattering length and effective
range that determine the asymptotic behaviour of $\lambda$.

At short distances the behaviour is different. The realistic LM2M2 model
with a strong repulsive core produces a strongly repulsive eigenvalue. The
eigenvalue from the exponential potential converges to $\lambda (0)=0$
as it does for all potentials which diverge slower than $r^{-2}$ at the
origin \cite{review}. The eigenvalue from the zero-range model converges
to $\lambda(0)=-4$ according to (\ref{extbc}). This is precisely
sufficient to eliminate the Thomas effect.

\begin{figure}[t]
\begin{center}
\input{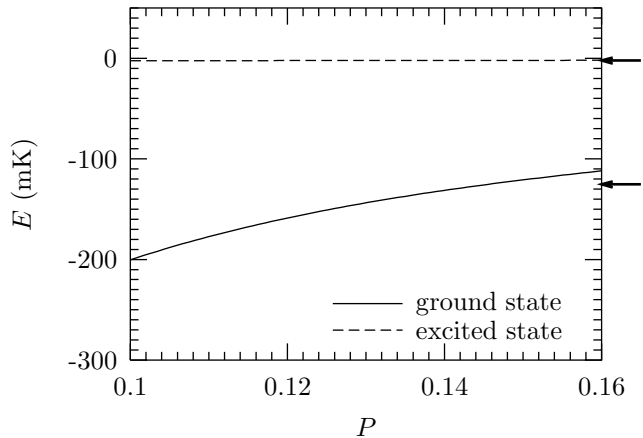}
\end{center}
\caption{The energies of the ground and the excited state of
$^{4}$He-trimer as function of the $P$ parameter of the extended
zero-range model.  The arrows indicate the results of the realistic
LM2M2 model \protect\cite{esbenJPB}.}
\end{figure}

Although the zero-range model gives a stronger attraction at small
distances the correct large distance behaviour and a good overall
agreement make it a solid alternative to finite-range potentials. One
would expect that a weakly bound three-body state, not sensitive to the
short-range details of the potential, should be reasonably well described
by the zero-range model while one would expect some over-binding for
strongly bound states.

\paragraph{Bound states.} The extended zero-range model correctly predicts
the number of $^4$He-trimer bound states -- two states -- for a large
variation of the $P$ parameter, see Fig.~2. The energy of the weakly
bound excited state, predicted rather accurately, is largely independent
of the $P$-parameter since it mostly resides in the outer region which
is determined exclusively by the scattering length and effective range.
The stronger bound ground state is more sensitive to the inner part of
the effective potential, and therefore to the $P$-parameter.  On average
the zero-range model gives a description similar to finite-range models,
see Table~1.

\begin{table}[tbh] \label{he4}
\caption{ The bound state energies of the helium trimers
$^4$He$_3$ and $^4$He$_2{}^3$He for finite-range potential models
\protect\cite{esbenJPB}, and for the zero-range model with $P$=0.13.
The $^4$He-$^4$He scattering length is $a$=-189.054~au and effective range
$R$=13.843~au. For $^4$He-$^3$He system $a$=33.261~au, $R$=18.564~au. The
mass of $^3$He is $m$($^3$He)=3.016026.  For the gaussian, exponential
and zero-range models the shown energies are calculated within the
adiabatic (one-channel) approximation (\protect\ref{adia}). For these
simple potentials the realtive accuracy of the adiabatic approximation
is better than 1\%.  The LM2M2 energies are obtained with the full expansion
(\protect\ref{expan}). }

\begin{center}
\begin{tabular}{|l||c|c|c|}
\hline
Potential & $E_0$($^4$He$_3$) (mK) & $E_1$($^4$He$_3$) (mK) &
		$E_0$($^4$He$_2{}^3$He) (mK) \\ 
\hline
\hline
LM2M2       & -125.2 & -2.269 & -13.66\\ 
Gaussian    & -150.2 & -2.462 & -18.41\\ 
Exponential & -173.9 & -2.714 & -24.27\\ 
Zero-range  & -143.7 & -2.21  & -34.0 \\
\hline
\end{tabular}
\end{center}

\end{table}

\paragraph{Radial functions.} Fig.~3 compares the radial functions from
different models.  Due to the softness of the zero-range model at short
distances the ground state wave-function is shifted to the left in
comparison to the exponential and LM2M2 potentials. However, it is still
rather similar to those obtained from the finite-range models. The
wave-functions for the excited state are roughly identical for all models
since the spatially extended weakly bound states are less sensitive to
the individual features of the underlying potential model.

\begin{figure}[t]
\begin{center} \input{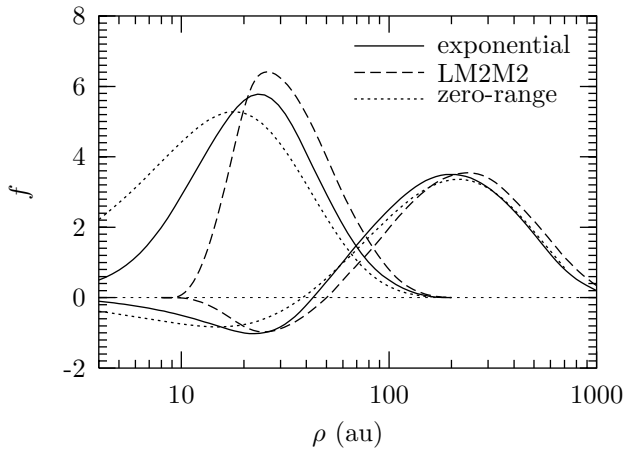} \end{center}
\caption{The arbitrarily normalized radial wave-functions $f$ as function
of $\rho$ for the ground and the excited states of the $^4$He trimer for
the potential models from Fig.~1.}
\end{figure}

\paragraph{Non-identical particles.} Another bound atomic trimer
$^4$He$_2{}^3$He is obtained by substituting $^3$He for one of the
$^4$He atoms.  The scattering length in the subsystem $^3$He-$^4$He is
$a$=33.261~au and the effective range $R$=18.564~au.  In order to obtain
$\nu(\rho)$ for the system of non-identical particles we have to solve
the general equation (\ref{det}) .

Although the $P$ parameter for the $^3$He-$^4$H subsystem should generally
speaking be also different, we choose the same value $P$=0.13 for the
sake of simplicity.  With these parameters the zero-range model correctly
predicts that there is only one bound state in this system. The binding
is somewhat larger than for finite-range potentials but within the same
range of accuracy, see Table~1. Again, we did not make any attempt to
fit the binding energy by varying $P$.

\section{Conclusion}

The zero-range potential is a very useful form of effective interaction
whose applications to three-body systems are, however, severely hampered
by the Thomas collapse.  We propose a coordinate space regularization
of the zero-range potential which leads to a removal of the Thomas
collapse. The new model on one hand retains all the simplicity of the
zero-range potential and on the other hand provides a fully regularized
solution for the three-body system.  Compared to finite-range potentials
the computational load is greatly reduced and amounts to solving a
transcendental equation for the effective potential and subsequent
ordinary differential equation for the radial wave-function.  We apply the
proposed model to atomic helium trimers and show that it works
well and produces results comparable to finite-range models.

\appendix

\section{Hyper-spheric coordinates}

If $m_{i}$ and ${\bf r}_{i}$ refer to the $i$-th particle then the
hyper-radius $\rho $ and the hyper-angles $\alpha _{i}$ are defined in terms
of the Jacobi coordinates ${\bf x}_{i}$ and ${\bf y}_{i}$ as \cite{RR} 
\begin{eqnarray} \label{hyp} 
{\bf x}_{i} =\sqrt{\mu _{i}}({\bf r}_{j}-{\bf r}_{k})\;,\;
{\bf y}_{i}= \sqrt{\mu _{jk}}\left( {\bf r}_{i}-\frac{m_{j}{\bf r}_{j}+m_{k}
{\bf r}_{k}}{m_{j}+m_{k}} \right)\;, \nonumber\\
\mu _{i} =\frac{1}{m}\frac{m_{j}m_{k}}{m_{j}+m_{k}}\;,\;
\mu _{jk}=\frac{1}{ m}\frac{m_{i}(m_{j}+m_{k})}{m_{i}+m_{j}+m_{k}} \\
\rho \sin (\alpha _{i}) =x_{i}\;,\;\;\rho \cos (\alpha _{i})=y_{i}\;, 
\nonumber
\end{eqnarray}
where $\{i,j,k\}$ is a cyclic permutation of \{1,2,3\} and $m$ is an
arbitrary mass. The set of angles $\Omega _{i}$ consists of the hyper-angle $
\alpha _{i}$ and the four angles ${\bf x}_{i}/|{\bf x}_{i}|$ and ${\bf y}
_{i}/|{\bf y}_{i}|$. The kinetic energy operator $T$ is then given as 
\begin{eqnarray}
&&T=T_{\rho }+\frac{\hbar ^{2}}{2m\rho ^{2}}\Lambda\;,\;
T_{\rho }=-\frac{\hbar ^{2}}{2m}\left( \rho ^{-5/2}\frac{\partial ^{2}}{
\partial \rho ^{2}}\rho ^{5/2}-\frac{1}{\rho ^{2}}\frac{15}{4}\right) \;,
\label{def} \\
&&\Lambda=-\frac{1}{\sin (2\alpha _{i})}\frac{\partial ^{2}}{\partial
\alpha _{i}^{2}}\sin (2\alpha _{i})-4+\frac{l_{x_{i}}^{2}}{\sin ^{2}(\alpha
_{i})}+\frac{l_{y_{i}}^{2}}{\cos ^{2}(\alpha _{i})}\;,  \nonumber
\end{eqnarray}
where ${\bf l}_{x_{i}}$ and ${\bf l}_{y_{i}}$ are the angular momentum
operators related to ${\bf x}_{i}$ and ${\bf y}_{i}$.

\end{document}